\documentstyle[epsfig,psfig]{mn2e}

\title{Cold Dust in (Some) High z Supernova Host Galaxies}
\author[D.L. Clements et al.]
       {D.L. Clements$^1$, D. Farrah$^2$, M. Rowan-Robinson$^1$, J. Afonso$^3$,R. Priddey$^4$, M. Fox$^1$\\
        $^1$Imperial College, London, Blackett Lab, Prince Consort Road, London SW7 2BW, UK\\
        $^2$IPAC, Caltech, 770 S. Wilson Ave., Mailstop 100-22, Pasadena, CA 91125, USA\\
        $^3$CAAUL, Universidade de Lisboa, Faculdade de Ci\^{e}ncias,Observat\'{o}rio Astron\'{o}mico de Lisboa, Tapada da Ajuda, 1349-018 Lisboa\\
	$^4$Department of Physics, Astronomy and Mathematics, University of Hertfordshire, Hatfield, Herts., UK}
\date{}

\pagerange{\pageref{firstpage}--\pageref{lastpage}}
\pubyear{}

\begin{document}

\maketitle

\label{firstpage}

\begin{abstract}
We present deep submillimetre photometry for 14 galaxies at z=0.5 that
are hosts of type 1a supernovae, with the aim of examining the
evolution of dust mass and extinction in normal galaxies.  We combine these
results with our previous observations of 17 z$\sim$0.5 SN1a hosts to
look for any evolution in the dust content of normal galaxies between
z=0 and z=0.5. The average observed frame 850$\mu$m flux of SN1a hosts
in the full sample, excluding 2 bright individually detected objects,
is 0.44 $\pm$ 0.22 mJy. This flux level is consistent with there being
little or no evolution in the dust content, or optical extinction, of normal
galaxies from z=0 to z=0.5. One galaxy, the host of
SN1996cf, is detected individually, and we also present a deep HST
STIS image for this object. It appears to be an edge on disk system,
similar to the submm bright host of SN1997ey.  We thus examine the dust properties of these
and one other individually detected object. 450-to-850 $\mu$m flux ratios and
limits suggest that the dust in the two brightest submm sources,
SN1996cf and SN1997ey, is cold, T $\sim$20K, implying that they
contain a substantial mass of dust $\sim10^9 M_{\odot}$. The presence
of two bright (F$_{850} >$7mJy) submm sources at z$\sim$0.5 in a sample
of ostensibly normal galaxies is surprising, and has important
implications.  It supports the idea that a substantial part of the
Cosmic Infrared Background (CIB) may be produced at z$<1$, while also
suggesting that 'foreground' objects such as these may be a
significant 'contaminant' in submm surveys. Finally, we examine
the overall submm luminosity distribution at z=0.5 implied by our
results, and conclude that either there is substantial evolution in
the submm luminosity function from z=0 to 0.5, or our submm detected
sources are somehow not representative of the bulk of galaxies at this
redshift.
\end{abstract}


\section{Introduction}

Two of the major new insights of the 1990s into cosmology and the
history of galaxy formation were the discovery of the Cosmic Infrared
Background (CIB: Puget et al., 1996; Fixsen et al., 1998) and the
measurement of a significant 'dark energy' term in the expansion of
the universe through the use of high redshift type 1a supernovae as
standard candles (Riess et al., 1998, HZT; Perlmutter et al., 1999,
SCP). The discovery of the CIB indicated that dust enshrouded star
formation is a significant aspect of the star formation history of the
universe, amounting to 50\% or more of all star formation (Gispert et
al., 2000). At the same time surveys with SCUBA (eg. Smail et al.,
1997; Hughes et al., 1998; Eales et al., 2000; Mortier et al., 2005)
and other submillimetre array detectors have begun to find the objects
that make up the CIB. These objects are largely interpreted as being
similar to local Ultraluminous Infrared Galaxies (ULIRGs), and are
thought to contain large dust masses at a temperature T$\sim$40K
(Blain et al., 1999) and to be the hosts of massive bursts of star
formation, $100-1000 M_{\odot} yr^{-1}$. However, substantial
uncertainties remain. Most of these submillimetre galaxies (SMGs) do
not have well determined redshifts (though see Chapman et al.,
2005). The degeneracy between temperature and redshift (Blain 1999)
thus means that their dust temperature is highly uncertain --- a low
dust temperature object at low redshift (z$\sim$0.5---1) looks much
the same as a high dust temperature object at high redshift (z$\sim$
2---3). Since luminosity is a strong power of temperature, typically
$T^6$ for a standard SMG spectral energy distribution (SED), this
leads to a considerable uncertainty in the derived luminosity and star
formation rates for those SMGs without measured redshifts. Indeed some
authors have suggested (Rowan-Robinson 2001, Kaviani et al. 2003, Efstathiou \&
Rowan-Robinson, 2003) that a substantial fraction of the SMG
population may be cooler and closer than originally thought, and
there is some observational evidence to back up this suggestion
(Chapman et al., 2003; Taylor et al in prep). There is currently
little known about cool dust in normal galaxies in the 0.5$<$z$<$1
range because it is quite difficuly to find a clean sample of such
objects. The SN1a host galaxies (see eg. Sullivan et al. (2003); Tonry et al. (2003)), though, provide an ideal selection
for such studies.

The discovery of the CIB also raises the overall question of the
evolution of the dust content of galaxies and its role in obscuring
starlight in these objects and reprocessing it into thermal dust
emission. Dust obscuration is known to have a significant effect
locally (Tresse \& Maddox, 1998), but the evolution of dust content
and obscuration is not well constrained. Chemical evolution models
(Calzetti \& Heckman, 1999; Pei, Fall \& Hauser, 1999) predict that
dust obscuration should peak at 2-3 times the current value at $1<z<2$
before gradually declining to higher redshifts. However, we currently
know little about the dust content of quiescent systems which make up
the bulk of the galaxy population at any epoch. The host galaxies of
high z supernovae, however, have not been selected on the basis of
active star formation, but have spectroscopic redshifts and abundant
ancillary data, including, for many, HST images and multicolour
optical and near infrared colours. They are thus an ideal sample with
which to study the role of dust in quiescent galaxies at cosmological
redshifts.

This paper is the second to result from our ongoing search for dust
emission in high z SN host galaxies. Our first paper (Farrah et al.,
2004, hereinafter Paper 1) presented the results of SCUBA submm
photometry of a sample of 17 z=0.5 SN1a host galaxies. We here present
additional observations of a further 14 galaxies, extending the sample
size to 31, and increasing the number of directly detected objects to
three. The paper is structured as follows. In section 2 we discuss the
observations and present our results. In section 3 we discuss the
sample as a whole and discuss the overall dust content of the
quiescent galaxy population at z=0.5. In section 4 we examine the
properties of the three galaxies individually detected in our
observations, while in section 5 we discuss the implications of our
results for the overall submillimetre luminosity distribution of
galaxies at z=0.5. We draw our conclusions in section 6. We assume
$\Omega_0 =1, \Omega_m = 0.3, \Omega_{\Lambda}=0.7$ and $H_0 = 70 kms^{-1}
Mpc^{-1}$ throughout this paper.

\section{Observations}

The observations described here extend the earlier work of Paper 1 on
the host galaxies of z$\sim$0.5 type 1a supernova host galaxies. As
with our Paper 1 observations, our targets were selected from the HZT
and the SCP supernova lists to be the host galaxies of type 1a
supernovae and to lie in a narrow ($\pm$ 0.1) range of redshifts
centred on z=0.5. This is to allow easy combination of submm fluxes
from the targets with no need for $k$-corrections to shift the fluxes
to correspond to the same rest-frame wavelength.

Observations were made of 14 high redshift supernova 1a host galaxies
at z$\sim$ 0.5 between July 2003
and January 2004 (see table 1). Observing conditions where
generally good, with $\tau_{850}$ ranging from 0.2 to 0.15 and
$\tau_{450}$ ranging from 1.3 to 0.85. These optical depth values were
determined by standard skydip observations at azimuths relevant to the
targets and are classified as grade 1 or 2 conditions by the JCMT.

The sources were observed with SCUBA (Holland et al., 1999) in
photometer mode using two bolometer chopping. This means that for each
channel, 850 or 450 $\mu$m, as well as being observed by the central
bolometer on the array, a second bolometer observes the source in the
reference position. An observed flux can be extracted from both these
bolometers which are then combined, using appropriate noise weighting,
to increase the sensitivity over that of a normal single
bolometer chopping observation by $\sim \sqrt 2$. This combination takes place at the
end of the data reduction process, once flat fielding, background
subtraction and calibration have been completed.

The data from SCUBA was reduced in a standard way using the SURF data
reduction package (Jenness \& Lightfoot, 2000). Regular pointing
observations were made to ensure pointing accuracy, and skydips were
also taken regularly. Flux calibration factors (FCFs) for the two
bolometers used to detect the sources were calculated
separately. Calibration sources used were CRL618 and Uranus. The
individual observations of a source with a given bolometer are
combined using a Kolmogorov-Smirnoff technique which masks out any
discordant points that remain after the data processing. No unusual
numbers of points were excluded in this analysis. When an object was
observed on more than one day, the fluxes measured in each separate
observation are combined using a variance weighted scheme.

Our results are presented in table 1. Only one source, the host galaxy
of SN1996cf, a z=0.57 SN1a, was detected, with a flux of 11$\pm$1.6
mJy at 850$\mu$m. This source was only 'detected' with 1.6$\sigma$
significance at 450$\mu$m, with a flux of 25$\pm$15 mJy.

For the host galaxy of SN1996cf, we obtained Hubble Space Telescope
(HST) imaging from the HST data archive. Observations were taken using
the Space Telescope Imaging Spectrograph (STIS) in imaging mode using
a clear filter. The data consisted of three exposures, each of 300
seconds, taken with a 10 pixel offset between successive exposures to
facilitate the subtraction of cosmic ray events. The data were
reduced, and combined into a single image, using the IRAF reduction
package {\it calstis}. The image of the host galaxy, presented in Fig. 1, has had the low surface brightness host enhanced by the application of a 2x2 pixel boxcar filter.

\begin{figure*}
\epsfig{file=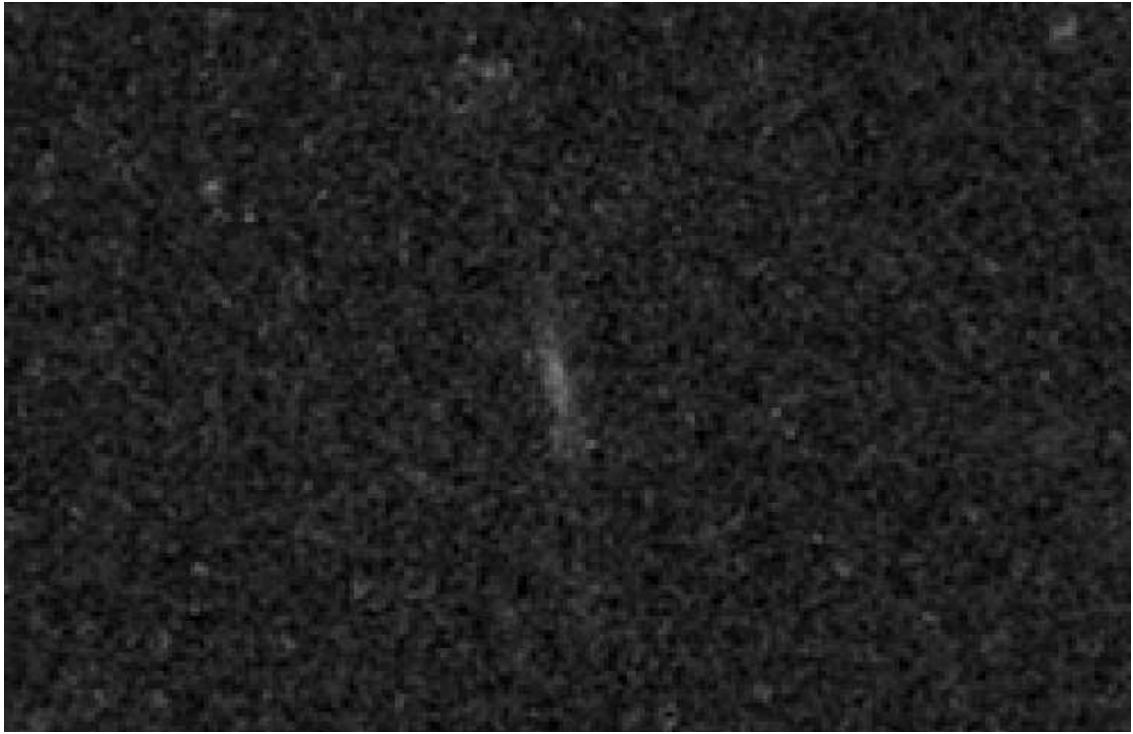,width=15cm}
\caption{STIS clear CCD image of the host galaxy of SN1996cf.}.
The pixel size in this the standard STIS CCD resolution of 0.05". The Image is 15"x11" in size, ie. 300x220 pixels.
\end{figure*}

\begin{table}
\begin{tabular}{cccc}\hline
Name&z&Date Observed&F$_{850}$(mJy)\\ \hline
SN1995az&0.45&9 July 2003&1.4$\pm$0.9\\
SN1997ax&0.615&26 Dec 2003&1.8$\pm$0.8\\
SN1997K&0.592&16 Dec 2003, 9 Jan 2004&-0.3 $\pm$0.9\\
SN1997H&0.526&8, 10 Jan 2004&0.7 $\pm$0.9\\
SN1997eq&0.538&8 Jan 2004&1.5$\pm$1.2\\
SN1997j&0.619&8 Jan 2004&0.4 $\pm$0.9\\
SN1997af&0.579&9, 10 Jan 2004&0.0$\pm$0.8\\
SN2000ec&0.47&10 Jan 2004&-2.7$\pm$1.4\\
SN1996au&0.52&10 Jan 2004&-3.0$\pm$1.4\\
SN1997es&0.65&10 Jan 2004&-0.7$\pm$1.4\\
SN1996cf$^*$&0.57&10 Jan 2004&11.0$\pm$1.6\\
SN1996ci&0.5&11, 12 Jan 2004&-0.7$\pm$0.8\\
SN1997aj&0.581&12 Jan 2004&-1.8$\pm$1.3\\
SN1996I&0.57&12 Jan 2004&-0.6$\pm$1.2\\
\end{tabular}
\caption{Results of submm observations of SN1a host galaxies.}
$^*$SN1996cf 450$\mu$m flux 25$\pm$15 mJy.
\end{table}

\section{The Evolution of Obscuration in Normal Galaxies}

The evolution of dust content and dust obscuration in normal galaxies
can be examined by looking at the average properties of the sample we
have observed. We follow the approach discussed in Paper 1, but apply
it to the full sample of 31 galaxies --- 17 from Paper 1 and 14 from
the current paper. We exclude 2 objects from this combination, the two
objects whose fluxes are strongly detected --- SN1996cf (this paper)
and SN1997ey (Paper 1). The fluxes are combined using an inverse
variance weighting scheme. The average observed frame
850$\mu$m flux from type 1a supernova host galaxies calculated in this
way is 0.44 $\pm$ 0.22 mJy. This is less than the mean flux of
1.01$\pm$0.33 mJy obtained in Paper 1, but still statistically
consistent (at the 1.5$\sigma$ level) with the previous result. If we
attempt to split the SN host galaxy sample into spirals (and
irregulars) and ellipticals, based on the available data on the host
galaxy morphologies (largely HST images), whilst still excluding the
two strong detections, we find a similar average flux for the two
classes of object (0.40 $\pm$0.29 for spirals/irregulars, and 0.69
$\pm$ 0.46 for ellipticals).

By consideration of the local SLUGs 850$\mu$m luminosity function (Dunne et al., 2000 - though see also Vlahakis et al., 2005) combined with appropriate {\em k} corrections from the cirrus model of Efstathiou \& Rowan-Robinson (2003), and the assumption that the submillimetre emission from ellipticals is negligible, Paper 1 predicted that the mean observed frame 850$\mu$m flux of a
z=0.5 galaxy would be 0.56 $\pm$0.1 mJy if there is no evolution in
the amount of dust in normal galaxies from z=0 to z=0.5. Our result is
thus consistent with there being at most only moderate evolution in
the dust content of normal galaxies from z=0 to z=0.5. Paper 1 found a
moderate increase in the dust content of 25\%--135\% over this
redshift range. The new expanded SN host sample thus mildly
contradicts this result (1.5 $\sigma$), though the errors permit an increase of up to
20\% or a 100\% decrease in the dust content of the population as a
whole. The mean A$_V$ for galaxies at z=0.5 would thus be statistically unchanged
from the local value of A$_V$ = 0.31 derived by Rowan-Robinson
(2003).This is in contrast with models of opacity evolution
from Calzetti \& Heckman (1999), which predict an increase in
extinction of A$_V \sim 0.15$ by z=0.15, and with chemical evolution
models (Pei et al., 1999). Neither of these models predict a
particularly significant evolution in extinction or dust mass in
galaxies over the period to z=0.5, so our current results cannot be
interpreted as a rigorous test of these models.  To achieve this, a
larger sample and more sensitive observations would be needed,
allowing more than statistical non-detections of the population,
and/or the greater lever arm that would be obtained by examining the
dust content of normal galaxes out to z=1. 

All of these considerations apply to the population as a
whole. Throughout this discussion we have excluded the two strong
submm detections in our sample from consideration because the two
objects appear to have radically different submm luminosity to the bulk of the population. We now consider the nature of these two objects in detail.

\section{The Properties of the Detected Galaxies}

Whilst we have failed to detect the mean submm emission of the population
of z$\sim$0.5 SN1a host galaxies, we have managed to detect three of
these objects individually. The SN magnitudes for these objects lie in typical positions for z$\sim$0.5 SN1a objects in the magnitude-redshift diagram (eg. Perlmutter et al., 1999). One of our sources, SN2000eh, was only
marginally detected (3$\sigma$) at 850$\mu$m, but two others, SN1997ey and SN1996cf
were detected surprisingly strongly. Their observational and derived
properties are given in Table 2. It can clearly be seen that their
rest-frame 850$\mu$m luminosities are comparable with those of local
ULIRGs such as Arp220 and Mrk231 (Dunne et al., 2000; Farrah et al., 2003). However, there
is no evidence for starburst or AGN activity in these objects on the
basis of optical spectra obtained by the high redshift supernova
teams (HZT \& SCP teams, private communication). HST imaging of the two bright submm sources also shows them to
be morphologically different from local ULIRGs and from those SMGs for
which imaging is currently available. Both of our bright submm sources
(see fig. 1 for 1996cf and Paper 1 for 1997ey) appear to be faint
quiescent edge-on disk galaxies. Local ULIRGs, in contrast, are almost
universally disturbed systems (eg. Clements et al., 1996; Surace et al., 1998; Farrah et al., 2001), with double
nuclei, tidal tails and other signs of merging activity. The SMG
population, as revealed by blank field SCUBA surveys, also appears to
be made up of such disturbed objects (eg. Clements et al 2004; Smail
et al., 2004; Chapman et al., 2003b), though it should be noted that
to date no flux limited SMG sample has been completely imaged.

The spectral energy distributions and dust temperatures of these
objects are difficult to determine since they are detected in only one
or two submm bands. The best we can attempt is to set constraints on
the temperature and SEDs by examining the 450/850 flux ratio and its
limits. These targets were all observed in excellent conditions, so
calibration errors should be smaller than or comparable to the
observational errors on the fluxes (see eg. Dunne \& Eales, 2001).
One source is clearly detected at 450$\mu$m, SN1997ey, while a second,
1996cf, has a marginal 1.6 $\sigma$ signal. There is no useful 450$\mu$m
data available for SN2000eh since it was observed in poorer
conditions. Examination of the 450/850 flux ratio for 1997ey and
1996cf (see fig. 2) suggest cold dust temperatures $\sim$20K assuming
an emissivity index of $\beta$=1.3 (consistent with the single
temperature SEDs found for local galaxies by the SLUGS survey (Dunne
et al, 2000)) and still colder for the more generally accepted
emissivity index $\beta$=2 (see eg. Dunne \& Eales 2001). Assuming a
standard dust opacity of 0.077 m$^2$kg$^{-1}$ (eg. Dunne et al. 2000), these would correspond
to dust masses of 1.3$\times 10^9 M_{\odot}$ and 1.7 $\times 10^9
M_{\odot}$ respectively. In the absence of 450$\mu$m data, we assume
that SN2000eh has similar dust properties, yielding a lower dust mass
of 6.5 $\times 10^8 M_{\odot}$. These are substantial dust masses,
about a factor of 10 more dust than in the dustiest galaxy in the
original SLUGS survey, UGC9618 (Dunne et al., 2000). The low inferred
temperature, combined with the large observed 850$\mu$m flux, lead
inevitably to this large dust mass. If the cooler temperature
associated with a $\beta = 2$ SED is assumed, then the dust masses get
correspondingly larger.

\begin{figure*}
\epsfig{file=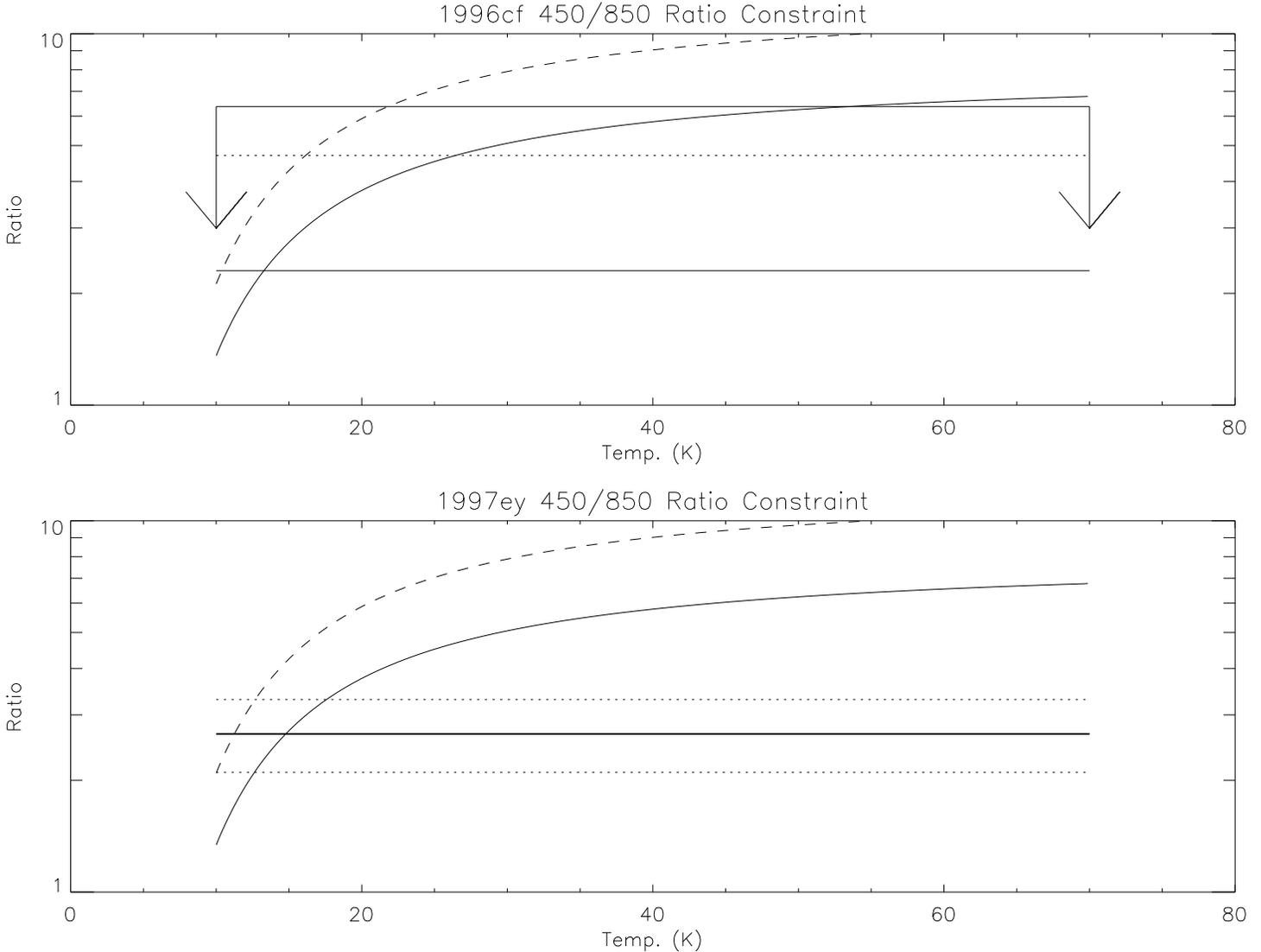,width=15cm,angle=90}
\caption{Comparison of Dust SED models to 450/850 ratio of strongly detected sources}
The curved lines are the predicted observed-frame 450/850 ratios for
thermal dust emission models with a $\beta$=1.3 emissivity (dashed
line) or $\beta$=2 emissivity (solid line). The horizontal lines are
the observed ratios. Solid lines show the observed values and, for
SN1996cf, the upper limit to this ratio using the formal 450$\mu$m
3$\sigma$ flux upper limit (designated with an arrow), while dashed
lies show the 1$\sigma$ range for this flux ratio. As can be seen,
SN1997ey appears to have a cold SED, with a temperature of 20K or
less, while it seems likely that SN1996cf's host is similar.
\end{figure*}

Dust as cold as we have found in these objects is not unknown in the
local universe (eg. the SED of NGC891 is dominated by dust at $\sim$
15K (Alton et al., 1998)) and has been suggested for higher redshift
objects on both theoretical grounds (Rowan-Robinson, 2001; Efstathiou
\& Rowan-Robinson 2003; Kaviani et al., 2003) and observational
grounds (Chapman et al., 2003; Taylor et al., 2005). The present
results are direct confirmation that cold galaxies with
large dust masses and submm luminosities do indeed exist at moderate
redshift, and could contribute to the CIB and to the sources seen in
deep far-IR surveys such FIRBACK (Puget et al., 1999) and deep submm
surveys. Indeed, objects such as these might prove to be a
foreground contaminant in deep large area submm surveys such as SHADES
(Mortier et al., 2005) and those now been planned for new facilties
and instruments such as SCUBA2 and the LMT (Large Millimetre
Telescope). The areal density for sources such as the two bright
galaxies discussed here was predicted in Paper 1 to be $\sim$100 per
sq. deg.  The results of this paper, which roughly doubles the sample
size and doubles the number of bright sources found, adds credence to
this number, which is comparable to the source densities found in deep
blank field submm surveys at these flux levels (eg. Scott et al.,
2002; Mortier et al., 2005). The presence of such foreground
interlopers might thus be a significant problem when attempting to
measure correlation functions for high redshift submm galaxies in the
absence of clear identifications and redshifts, especially as they are unlikely to be bright in the radio.

\begin{table*}
\begin{tabular}{ccccccc} \hline
Name&RA&DEC&z&F$_{450}$ mJy&F$_{850}$ mJy&L$_{850}$ W m$^{-2}$sr$^{-1}$\\ \hline
SN1997ey&04 56 58.2&-02 37 37&0.58&20.8$\pm$3.5&7.8$\pm$1.1&8.7$\times 10^{22}$\\
SN1996cf&10 48 50.9&00 03 30&0.57&25$\pm$15&11.0$\pm$1.6&1.2$\times 10^{23}$\\
SN2000eh&04 15 02.4&04 23 18&0.49&5.2$\pm$6.6&4.6$\pm$1.5&4.7$\times 10^{22}$\\
\end{tabular}
\caption{Detailed Observational Properties of the strongly detected sources.
For comnparison, Arp220, the nearest ULIRG, and Mrk231, the nearest
type 1 AGN-ULIRG, have 850$\mu$m luminosities of 4.0 $\times 10^{22} W
m^{-2}sr^{-1}$ and 1.0 $\times 10^{23} W m^{-2}sr^{-1}$ respectively.}
\end{table*}

\section{z=0.5 Submm Luminosity Distribution}

In principle, our sample of SN1a host galaxies should provide a
reasonably unbiased selection for evolved galaxies at z=0.5,
biased in selection only by the stellar mass. By examining the
850$\mu$m luminosity distribution of these objects, and comparing it
with that found for local objects, we can gain some insight into the
galaxy population at z=0.5. The local luminosity function that we use
is the SLUGS 850$\mu$m LF (Dunne et al 2000) since this is the best
currently available. This was found to be well fitted by a Schechter
function with L$_*$ = 8.3 $\times 10^{21} W m^{-2} sr^{-1}$, $\alpha =
-2.18$ and $\Phi_{*} = 2.9 \times 10^{-4} Mpc^{-3} dex^{-1}$ We cannot
apply this luminosity function directly to predict the numbers of
sources we should see, since we do not know the parent population
normalisation for the number of our target objects per
Mpc$^{-3}$. However, since all our objects are at the same redshift,
$\sim$0.5, we can look at the relative number of objects in each
luminosity bin and compare this to the low z expectation. We have only
a small number of detected galaxies, so must necessarily normalise the
expected luminosity distribution at the high end, where our detections
lie.

The Schechter function form of the LF used by Dunne et al. (2000) has a
very strong exponential rolloff at high luminosities. The detection of
a small number of very luminous objects, as obtained here, would
imply the presence of many more lower luminosity systems. We find 3
sources in the luminosity range 10$^{22.5}$ -- 10$^{23.5} W m^{-2}
sr^{-1}$. The Dunne et al. LF would predict that the next lowest dex
bin in the LF should contain 100 times more objects. These would have
fluxes in the range 0.3 -- 3 mJy at 850$\mu$m, and would thus be
detectable by us either individually or statistically. And yet we find
no evidence for such sources. Perhaps more significantly, sources as
luminous as SN1996cf should be extremely rare - so rare in fact that
we would not expect to see any at all given that we are only seeing 2
sources in the luminosity range 10$^{22.5}$ -- 10$^{23} W m^{-2}
sr^{-1}$. The assumption of a polynomial tail at high luminosities to
the 850$\mu$m LF, rather than an exponential fall off, as suggested
for the 60$\mu$m LF (Sanders \& Mirabel, 1996) would ameliorate this problem
to an extent, but would not provide a complete explaination for the
numbers of high luminosity sources we are seeing, or the absence of
sources at somewhat lower luminosities that would still be detectable
directly or statistically.

Several possible explainations are possible:

\begin{itemize}

\item The Luminosity Function has evolved from z=0 to z=0.5

At face value, the most obvious explaination for the discrepencies in
the luminosity distributions would be evolution in the luminosity
function. This would not be one of the standard density or luminosity
evolution since this would result in a many more moderate
flux/luminosity detections than we see. It would also contradict the
absence of evolution in the opacity that we find for the sources
undetected individually. Instead, some kind of evolution that only
affects the high luminosity objects would be needed ie. some kind of
luminosity-dependent luminosity or density evolution. This could lead
to a bimodal luminosity function.

\item The Local Luminosity Function is incorrect as a result
of missing objects such as those we have detected

While the SLUGS survey is currently the best determination of the
local 850$\mu$m luminosity function, it consists of less than 200
galaxies. It is thus entirely possible that classes of object are
missing from it. If a class of cold dust temperature, high submm
luminosity source existed, they might very well be absent from
SLUGS. At the low dust temperatures of our galaxies, $\sim$ 20K, such
an object at z $\sim$0.1 would have a 60$\mu$m flux $\sim$ 70mJy, and
so would not have appeared in the IRAS catalogs and would not have
been targeted by SLUGS or any other submm followup. The corresponding
100$\mu$m flux would be $\sim$ 1 Jy, since this is reaching towards
the peak of the SED. Such a cold source might be written off as likely
cirrus contamination and thus be absent from the main IRAS galaxy
catalogs. Its 850$\mu$m flux would be about 100mJy, comparable to the
60$\mu$m flux. At a redshift of 0.05 the fluxes would be 4 times
larger and so such a source would still be only marginally
detected by IRAS at 60$\mu$m. If they exist, such objects would best be
detected by large area, possibly all sky, shallow surveys with
SCUBA2, or in the Planck all sky surveys.

\item The bright targets are not members of the same population as the
rest of our targets

Our application of the low z luminosity function is based on the
assumption that all of our target SN host galaxies come from the same
parent population of established galaxies with large stellar
populations. This might not be the case. If there was an additional
population of SN host galaxies with large submm luminosities not arising from quiescent dust heating, making
up $\sim$10\% of SN hosts, then it would not be correct to extrapolate
the submm properties of the bulk of the SN hosts from a few objects
coming from a separate population. How might such a situation arise?
The obvious class of objects with large submm luminosities are
galaxies hosting a substantial burst of star formation, though the low
dust temperature in our bright submm objects and their quiescent
morphology might argue against this. However, starburst galaxies are
not the obvious hosts of type 1a supernovae. They are, though, the
likely hosts of other classes of supernova. If some fraction of the
SN1a's in the surveys are actually misclassified core collapse
supernovae in actively starforming, dusty hosts - they might
perhaps be type 1c SNe, which have masqueraded as type 1a's in the past
(Kotak, private communication; see also Homeier, 2004) - then
this could provide an explanation. There
would, though, be consequences for the SN1a survey results if there is
this level of contamination by non-1a's. Core collapse supernovae are
typically of lower luminosity than type 1as. Type 1b and c SNe, viewed
as a class, have absolute magnitudes about 1.24 magnitudes fainter
than type 1a SNe (Richardson et al., 2002). If there is indeed a
$\sim$ 10\% contamination rate of type 1as by type 1cs in the surveys,
then this could produce an overall dimming of the mean brightness by
0.24 magnitudes. This is comparable to the dimming at z$\sim$0.5
attributed to the cosmological constant or dark energy term
(Perlmutter et al., 1999; Reiss et al., 1998), and thus might have a bearing on the
reality or value of this cosmological component. The redshift
evolution in the dimming of SN1as attributed to dark energy at higher
redshifts would not easily be mimicked by interloping type
1c SNe, while the absence of starburst spectral features in these objects
is an additional problem for this explanation.

\end{itemize}

At this stage it is not possible to choose between these scenarios
since the sample of such objects is still very small. Each possible
explaination has its own significant implications for astrophysics, so
it is important that this strange population is followed up in
more detail, and that steps are taken to expand the sample size. At
this stage the only way to find more such objects seems to be to
observe additional SN1a host galaxies, in which they are represented
at $\sim$10\%. This is not a particularly efficient method of
discovery. Extension of our SN1a host studies to z=1 might prove
beneficial, and could test whether these objects are the result of
evolution. Aside from that, the best hope for uncovering more of this
class might be the cosmological submm surveys being proposed for
SCUBA2. Furthermore, a very large area shallow SCUBA2 survey, possibly
covering the whole accessible sky, would be one way to test the
hypothesis that cold galaxies are a feature of the general galaxy
population at all redshifts by looking for their local
equivalents. Conversely, detailed study of our existing submm bright
targets, in search of signs of recent star formation, might be the
best way of testing our third explaination, and thus to remove any
lingering doubts about the classification of their supernovae. The
absence of clear starburst indications or of morphological disturbance
in existing data probably makes this final explanation the least
favoured of the three we have proposed.

\section{Conclusions}

We have extended our search for dust in the host galaxies of type 1a
supernovae at z$\sim$ 0.5. We combine our new observations with those
of Paper 1, and find that there is no evidence for a substantial
increase in optical obscuration in the galaxy population between z=0
and z=0.5. We have, though, detected strong submm emission from one
object, in addition to the one strong detection and one weaker
detection from Paper 1. HST imaging of the two bright submm sources
suggests they are quiescent, edge on disk galaxies, in contrast to the
disturbed and interacting SMGs, found in blank field submm surveys,
and local ULIRGs. We
examine the dust properties of these sources, and conclude that they
are ultra-luminous submm sources, with large dust masses, but with
cold dust SEDs, T$\sim$20K. We also examine the luminosity
distribution of the SN1a host galaxies as a class, and compare it to
the local SLUGS submm luminosity function. We conclude that the z=0.5
luminosity distribution seems bimodal, with an absence of sources
within a factor of a few of our bright objects. Several scenarios
could explain this effect, including some forms of luminosity function
evolution, the absence of sources such as these from local submm
surveys, or problems with the classification of $\sim$10\% of SNe in
the type 1a SN surveys. We propose various ways in which these
hyptheses could be tested by future observations.
\\~\\
{\bf Acknowledgements}
\\~\\
The James Clerk Maxwell Telescope is operated by The Joint Astronomy Centre on behalf of the Particle Physics and Astronomy Research Council of the United Kingdom, the Netherlands Organisation for Scientific Research, and the National Research Council of Canada. DLC acknowledges funding from PPARC. JA gratefully
acknowledges the support from the Science and
Technology Foundation (FCT, Portugal) through
the research grant POCTI-FNU-43805-2001. The authors would like to thank the staff at the JCMT for their usual excellent support work, and R. Kotak for useful discussions. The research described in this paper was carried out, in part, by the Jet
Propulsion Laboratory, California Institute of Technology, and was
sponsored by the National Aeronautics and Space Administration.
This publication is also based in part on observations made with the NASA/ESA Hubble Space Telescope, obtained from the Data Archive at the Space Telescope Science Institute, which is operated by the Association of Universities for Research in Astronomy, Inc., under NASA contract NASÊ5-26555.

\end{document}